\documentstyle[sprocl]{article} 
  \def\bea{\begin{eqnarray}}
\def\eea{\end{eqnarray}}
  \newcommand\beq{\begin{equation}}
 \newcommand\eeq{\end{equation}} \newcommand\beqn{\begin{eqnarray}}
 \newcommand\eeqn{\end{eqnarray}}

\begin{document}

\title{Energy Dependence of the Pomeron
Spin-Flip}

\author{B.Z.~Kopeliovich$^{1,2}$ and
B.~Povh$^1$}

\address{$^1$~Max-Planck-Institut f\"ur
Kernphysik\\
Postfach 30980, 69029 Heidelberg,
Germany}

\address{$^2$~Joint Institute for Nuclear
Research\\
141980 Moscow Region, Dubna,
Russia}


\maketitle\abstracts{There is no theoretical reason
to
think that the spin-flip component of the Pomeron is
zero.
One can measure the spin-flip part using
Coulomb-nuclear
interference (CNI).  Perturbative QCD calculations show
that
the spin-flip component is sensitive to the smallest
quark
separation in the proton, while the non-flip part probes
the
largest separation.  According to HERA results on the
proton
structure function at very low $x$ the energy dependence
of
the cross-section correlates with the size of the
color
dipole.  Analysing the data from HERA we predict that
the
ratio of the spin-flip to non-flip amplitude grows
with
energy as $r(s)\propto
(1/x)^{0.1-0.2}$, violating Regge factorization 
of the Pomeron.}

\section*{How to measure the Pomeron spin-flip?}

The Pomeron contribution to the elastic scattering
amplitude
of a spin $1/2$ particle has a
form \cite{bgl,thepaper},

\beq
f^P(s,t)=f^P_0(s,t)\,\left[1+i\,\frac{\sqrt{-t}}
{m_N}\ \vec\sigma\cdot\vec n\ r(s,t)\right]
\label{1}
\eeq
The function $r(s,t)$ characterises the Pomeron
spin-flip
to non-flip ratio. In the case of $NN$ elastic
scattering
$r$ can be expressed through the spin amplitude
in standard notations \cite{bgl,thepaper},
\beq
r=\frac{2m_N}{\sqrt{-t}}\,
\frac{\Phi_5}{{\rm Im}(\Phi_1+\Phi_3)}
\label{2}
\eeq

If the Pomeron is a Regge pole (factorization holds),
the
spin-flip and
non-flip
amplitudes have the same phase, i.e. $r$ is {\sl
pure
imaginary}. Either this is true, or ${\rm Re}\,r\ll
1$. Indeed, a real part of $r$ would lead to a polarisation
in the
elastic amplitude due to ''self-interference'' of the two
components
of the
Pomeron,

\beq
A_N^{pp}(t)=\frac{\sqrt{-t}}{m_N}\
\frac{4{\rm Re}\,r(t)}{1+|r|^2\,|t|\,/m_N^2}\
,
\label{3}
\eeq
which is measured to be less that $1\%$ at high
energies.
On the other hand, even if $|r|$ is quite large,
the
polarisation may be small provided that
factorization
approximately holds. As soon as the polarisation is
insensitive
to the Pomeron spin-flip, it makes it difficult to
measure
$r$ in elastic hadronic scattering.

A unique way to measure $r$ is to study the polarisation
effects due
to interference of electromagnetic and hadronic
amplitudes (CNI - Coulomb-Nuclear Interference)~\cite{kz}. 
The corresponding polarisation in
$pp$
elastic scattering reads~\cite{kl,bgl}
\beq
A^{pp}_N(s,t)=
A^{pp}_N(t_p)\,
\frac{4y^{3\over 2}}{3y^2+1}\
,
\label{4}
\eeq
where
$y=|t|/t_p$,
\beq
t_p(s)=\frac{8\pi\sqrt{3}\alpha_{em}}
{\sigma^{pp}_{tot}}\
,
\label{5}
\eeq
and
\beq
A^{pp}_N(t_p)=\frac{\sqrt{3t_p}}
{4m_p}\,(\mu-1)\
.
\label{6}
\eeq
Here $\mu-1=1.79$ is the anomalous magnetic moment of
the
proton.

It was assumed in~\cite{kl,bgl} that $Im\,r=0$,
otherwise
one should replace~\cite{kz}
\beq
(\mu-1) \Longrightarrow (\mu-1) -
2\,{\rm Im}\,r
\label{7}
\eeq
This provides a parallel shift of the
function
(\ref{4}) up or down dependent on the sign of
${\rm Im}\,r$.
Therefore, measurement of $A_N$ in the CNI region seems
to
be a perfect way to study
$r$.
First very crude measurements were performed by
the
E704 Collaboration~\cite{e704} at Fermilab with $200\
GeV$
polarised
proton beam. The results are in a very good agreement
with
the predictions (\ref{4})-(\ref{6}), but can be also used
to
establish soft bounds on the possible value of
${\rm Im}\,r$.
According to the analyses~\cite{abp,t} 
the data~\cite{e704}
demand
\beq
{\rm Im}\,r <
0.15 \pm 0.2
\label{8}
\eeq
Much more precise measurements are expected to be done
with
polarised proton beams at
RHIC.

Another available source of information about the
Pomeron
spin-flip is the data on polarisation in $\pi^{\pm}p$
elastic
scattering, which have a reasonable accuracy at
energies
$6-14\
GeV$.
The dominant contribution of the
$\rho$-Reggeon
and Pomeron interference cancels in the sum of
the
polarisations. The rest is due to Pomeron and
$f$-Reggeon
interference. Its value can be used for an upper bound
on
the Pomeron spin-flip (assuming that $f$ has no
spin-flip),
which was found to be less than $3\%$~\cite{k80,k97,thepaper}.

Theoretical attempts to estimate 
the value of $r$ led to nearly
the
same values.  In~\cite{itep} the two-pion exchange was
used
for the Pomeron-nucleon vertex.  It was found that
the intermediate 
nucleon and $\Delta$ essentially
cancel
each other in the iso-scalar t-channel exchange, but add
up
in the iso-vector channel.  They found ${\rm Im\, r}\approx 0.05$.

The two gluon model for the Pomeron was used in~\cite{kz,z}
to
evaluate the Pomeron spin-flip part. A quark-gluon
vertex
conserves helicity. This fact led to a wide spread
opinion
that the perturbative Pomeron has no spin-flip. This is
not,
however, true. The quark momenta are directed
differently
from the proton momentum due to transverse motion of
the
quarks. Therefore, the proton helicity is not equal to
the
sum of the quark helicities, and helicity conservation
for
the quarks does not mean the same for the
proton.

\section*{What distances in the proton are 
probed by the spin-flip
Pomeron?}

It was
found in~\cite{kz,z} that the quantity $r$ is extremely
sensitive
to the choice of the proton wave function. For a
symmetric
3-quark configuration all the contributions to the
proton
spin-flip amplitude cancel. Only if the proton wave function
is
dominated by an asymmetric quark-diquark configuration is
the
spin-flip amplitude nonzero~\cite{kz,z}. The
smaller the 
$qq$ separation in the diquark is, the larger is
the
spin-flip fraction. 
${\rm Im}\,r$
reaches nearly $10\%$ at small
$t$ if $r_D \approx 0.2\ fm$.

We conclude that the spin-flip part of the Pomeron
probes
the smallest distances in the proton. The smaller
the minimal
quark separation in the proton, the higher the virtuality of
the
gluons in the Pomeron has to be in order to resolve
this
small distance. At the same time, the non-flip part of
the
Pomeron probes the largest quark separation in the
proton
and remains nearly the same even if the diquark size tends
to
zero.

\section*{The energy dependence and HERA data}

One of the main discoveries at HERA is the $Q^2$
dependence of
the
effective Pomeron intercept: the higher is
$Q^2$,
{\it i.e.} the smaller is the size of
hadronic
fluctuations in the virtual photon, the more 
the virtual photoabsorption 
cross section grows with
energy.

As soon as the spin-flip and non-flip parts of the
Pomeron
probe different scales in the proton one should
expect
different energy dependences. To find the
correlation
between the effective Pomeron intercept and the size of
the
photon fluctuation we can use the factorized form of
proton
structure function~\cite{nz},
\beq
F_2^p(x,Q^2) \propto
\int\limits_{c/Q^2}^{c/\Lambda^2}
\frac{dr_T^2}{r_T^4}\,\sigma(r_T,x)
\label{9}
\eeq
The constant $c$ is of the order of
one,
so we fix
$c=1$.

The dipole cross section 
which depends on the transverse $q\bar q$ separation
$r_T$ and the Bjorken $x$ can be
parametrised as
\beq
\sigma(r_T,x) = \sigma(r_T)\left({1\over
x}\right)^{\Delta(r_T,x)}\ .
\label{10}
\eeq

The power $\Delta(r_T,x)$ can be interpreted as an
effective
Pomeron intercept, since it is nearly
$x$-independent.
It follows from (\ref{9}) that
\beq
\Delta(r_T=1/Q^2)=
\frac{d}{d\
ln(1/x)}\
ln\left[\frac{d}{d\,ln(Q^2)}\,F_2^p(x,Q^2)\right]
\label{11}
\eeq
We use the fit~\cite{kp} to the proton structure
function
at $Q^2>1\ GeV^2$. The result for $\Delta(r_T)$ is shown
in
fig.~1 for few values of $x$.

\begin{figure}[tbh]
\includegraphics{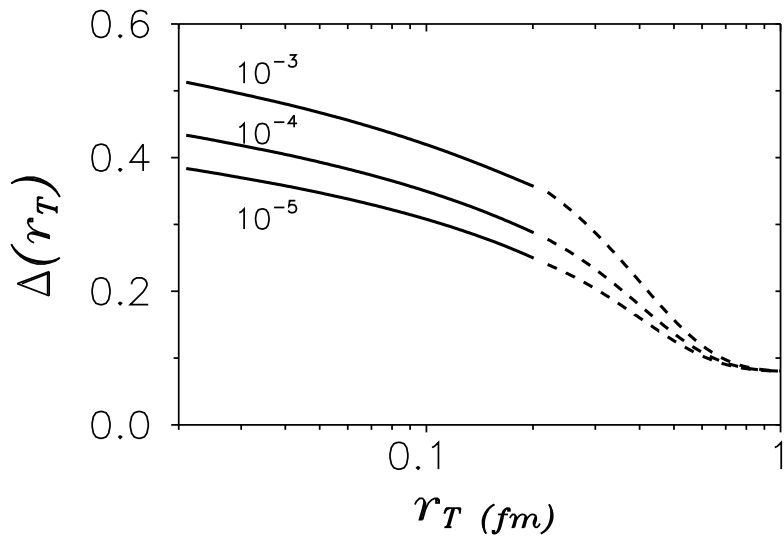}
\begin{center}
\vspace{7.5cm}
\parbox{10cm}
 {\caption[Delta]
 {\it The effective Pomeron intercept as function of the dipole size
as found using (\ref{11}) and HERA data (solid curves) at 
different values of $x$ shown at the curves. The dashed curves
are the guessed extrapolation to the soft hadronic limit $\Delta\approx
 0.08$.}
\label{fig1}}
\end{center}
\end{figure}

It extends only up
to
$r_T=0.2\,fm$ due to the restriction on $Q^2$. We
know,
however, that the $\Delta(r_T)$ keep
monotonically
decreasing at larger separations down to the
value
$\Delta\approx 0.08$, typical for soft hadronic
interactions. The dashed curves showing the interpolation
is just our guess.

The growth of $\Delta(r_T)$ down to small $r_T$ naturally
follows from the DGLAP evolution equations in the 
double-leading-log approximation (e.g. Refs.\cite{book}$^-$\cite{jetp}).
Decreasing the transverse separation $r_T$ in the color dipole
one increases the mean virtuality of the gluons which
have to resolve this small structure. Therefore, one 
gets a larger logarithmic factor ${\rm ln}(1/r_T)$ 
for each higher order correction in $\alpha_s$.
In the case of a spin-flip amplitude the mean gluon virtuality 
is related to the size of the diquark whose inner structure
is to be resolved. In the double-leading-log approximation 
this causes a steeper energy dependence
of the spin-flip amplitude in the same way as for the total 
cross section of a small-size color dipole.
Therefore, we can use the $r_T$-dependence plotted in Fig.~\ref{fig1} 
extracted from the analysis of data on $F_2(x,Q^2)$
to estimate the value of $\Delta$ for the spin-flip amplitude
assuming that $r_T$ is the diquark size.

We see from Fig.~\ref{fig1} that for the diquark 
size which is usually believed to
be $0.2-0.4\ fm$ \cite{diquark} the effective intercept of the
Pomeron
spin-flip ranges within $\Delta(r_T\approx0.2-0.3\,fm) 
\approx 0.2 - 0.3$. Therefore, the
fraction
of the spin-flip in the Pomeron increases with energy
as 
\beq
{\rm Im}\,r(s)
\propto
\left(\frac{s}{s_0}\right)^{0.1-0.2}
\label{12}
\eeq
Such a steep growth can be easily detected
with
polarised proton beams at RHIC whose energy range
covers (including fixed target experiments)
$50\,GeV^2 < s < 250000\,GeV^2$. The value of 
$r(s)$ more than doubles
in
this interval. This effect can be detected in
the
Coulomb-nuclear interference region in the
$pp2pp$
experiment planned at RHIC.

Summarising, we expect a rising with energy ratio
of spin-flip to non-flip components of the 
elastic proton-proton amplitude. This prediction is 
based on two observations: (i) Perturbative QCD calculations 
show that the spin-flip amplitude is sensitive to the smallest
transverse distance in the quark wave function of the proton.
Since the elastic amplitude is dominated by gluonic 
exchanges at high energies, one can conclude that the spin-flip
component probes the gluon distribution
in the proton at higher virtuality than the non-flip amplitude.
(ii) The proton structure function $F^p_2(x,Q^2)$ measured
at HERA rises with $1/x$. Data show that the higher is $Q^2$,
the steeper is the growth of $F^p_2(x,Q^2)$.

We conclude from (i) and (ii) that the spin-flip amplitude 
rises with energy faster than the non-flip one.
We expect this prediction to be tested in forthcoming polarisation
experiments at RHIC.
\\[.5cm]
{\bf Acknowledgements:}
We are thankful to Nigel Buttimore and Rudolph Hwa
who have read the paper and made
many valuable comments.

\end{document}